# Energy gap of topological surface states in proximity to a magnetic insulator

Jiashu Wang,[1]* Tianyi Wang,[1] Mykhaylo Ozerov,[2] Zhan Zhang,[3] Joaquin Bermejo-Ortiz,[4] Seul-Ki Bac,[1] Hoai Trinh,[1] Maksym Zhukovskyi,[5] Tatyana Orlova,[5] Haile Ambaye,[6] Jong Keum,[6,7] Louis-Anne de Vaulchier,[4] Yves Guldner,[4] Dmitry Smirnov,[2] Valeria Lauter,[6] Xinyu Liu,[1] Badih A. Assaf[1]*

[1] Department of Physics and Astronomy, University of Notre Dame, Notre Dame IN, 46556, USA
[2] National High Magnetic Fields Laboratory, Florida State University, Tallahassee, FL, 32310, USA
[3] X-ray Science Division, Advanced Photon Source, Argonne National Lab, Lemont, IL, 60439, USA
[4] Laboratoire de Physique de l'Ecole normale supérieure, ENS, Université PSL, CNRS, Sorbonne Université, 24 rue Lhomond 75005 Paris, France
[5] Notre Dame Integrated Imaging Facility, University of Notre Dame, Notre Dame IN, 46556, USA
[6] Neutron Scattering Division, Neutron Sciences Directorate, Oak Ridge National Laboratory, Oak Ridge, TN 37831, USA
[7] Center for Nanophase Materials Sciences, Physical Science Directorate, Oak Ridge National Laboratory, Oak Ridge, Tennessee 37831, USA
* Corresponding author. jwang39@nd.edu, bassaf@nd.edu

**Abstract.** Topological surface-states can acquire an energy gap when time-reversal symmetry is broken by interfacing with a magnetic insulator. This gap has yet to be measured. Such topological-magnetic insulator heterostructures can host a quantized anomalous Hall effect and can allow the control of the magnetic state of the insulator in a spintronic device. In this work, we observe the energy gap of topological surface-states in proximity to a magnetic insulator using magnetooptical Landau level spectroscopy. We measure $Pb_{1-x}Sn_xSe$–$EuSe$ heterostructures grown by molecular beam epitaxy exhibiting a record mobility and low Fermi energy. Through temperature dependent measurements and theoretical calculations, we show this gap is likely due to quantum confinement and conclude that the magnetic proximity effect is weak in this system. This weakness is disadvantageous for the realization of the quantum anomalous Hall effect, but favorable for spintronic devices which require the preservation of spin-momentum locking at the Fermi level.

**Introduction**

Topological insulators are materials that host gapless Dirac surface states due to a band inversion at high symmetry points in momentum space.[1,2] Charge carriers occupying these states have their spin degree of freedom locked to the momentum thus enabling spin-charge conversion. When these topological Dirac states are in proximity with magnetism, they acquire an energy gap due to broken time-reversal symmetry as illustrated in Fig. 1(a).[1,3,4] Topological insulator (TI) – magnetic insulator (MI) heterostructures are being pursued both for spintronic device applications [5,6,7,8,9] and to realize new quantum effects with broken time-reversal symmetry. The study of these heterostructures has thus been a consistent topic of interest both for applications and fundamental physics.

Fundamentally, the magnetic proximity effect is a route to introduce magnetism onto various states of matter without chemical doping.[1,3,10,11,12,13] It is being pursued in topological insulators for the realization of the quantized anomalous Hall effect (QAHE)[10] which requires topological surface state to have a large energy gap. In topological crystalline insulators (TCIs) which are topological materials with valley degenerate surface states protected by mirror symmetry, the QAHE effect has not yet been achieved. It is predicted to be large and tunable, owing to the valley degeneracy of TCIs enhancing the Chern number.[14]

Several experiments have studied the electrical and magnetic properties of heterostructures of topological and a magnetic insulator, but the electronic structure and energy gap of topological surface states in proximity with magnetism have yet to be measured.[3,15,16,17,18,19,20,21,22] A knowledge of the size of this gap is important as it provides as a metric of the strength of the magnetic proximity effect. The exact nature of spin-momentum locking in the presence of time reversal symmetry breaking also depends on the size of this energy gap.[23,24] Its measurement in the presence of a magnetic proximity with an insulator is thus important for spintronic devices. This measurement has not yet been done partly because most tools utilized to probe band structure are surface sensitive and cannot probe states located below a nanometer-thick MI layer. We will refer to states located at the TI/MI interface as topological interface states (TIS) in the remainder of this work. The mechanism by which the MI interacts with an underlying TIS remains unclear and controversial.[25,26] A measurement of the energy gap of the TIS can also shed light on this problem.

Here, we report a measurement of this energy gap $E_{TIS}$ when the TIS of a topological crystalline insulator are in proximity to an MI, in $Pb_{1-x}Sn_xSe$-EuSe heterostructure grown by molecular beam epitaxy (MBE) and sketched in Fig. 1(b). Magnetooptical Landau level spectroscopy allows us to measure $E_{TIS}$=14±6meV when the topological states are in proximity with magnetism. Polarized neutron reflectometry (PNR) and SQUID magnetometry measurements demonstrate that the alternating EuSe layers are in a fully saturated ferromagnetic state in proximity to the TIS at low temperature and high magnetic field. The origin of $E_{TIS}$ in our experiment is shown to be mainly due to quantum confinement, while magnetism and time-reversal symmetry breaking play a minor role despite the presence of the magnetic EuSe layer. Our work thus establishes a previously unknown upper limit on the TIS gap due to the magnetic proximity effect. Importantly, we find that in $Pb_{1-x}Sn_xSe$-EuSe, the Fermi energy is only 40meV above the middle of $E_{TIS}$ demonstrating the potential of topological crystalline insulators as high-quality alternatives for the realization of TI-MI devices.

Our material of choice $Pb_{1-x}Sn_xSe$ is a tunable topological crystalline insulator with degenerate surface states that are protected by mirror symmetry and time reversal symmetry.[27,28,29,30,14] It is ideal to test the proximity effect since it has a crystal structure identical to Eu-chalcogenides, commonly used in magnetic proximity effect devices.[3,31] Thus, It can be epitaxially matched to them without impacting sample quality.[32,33,34] This fact allows the proximity-induced gap of the TIS to be measured using Landau level spectroscopy. This method is a highly reliable tool for measuring narrow energy gaps[35,36,37,38] and for characterizing semimetals.[39,40,41] The $Pb_{1-x}Sn_xSe$-EuSe interface and our heterostructures also host a quasi-type-I band alignment with a band inversion and ensures that inversion symmetry is preserved. EuSe is however an antiferromagnet at zero field. But, by applying a large enough magnetic field along the growth axis of the heterostructure, needed for Landau level spectroscopy, we can saturate the magnetic moment of EuSe to a ferromagnetic state. In the configuration of our experiment, EuSe thus mimics a ferromagnetic insulator with a magnetization oriented out-of-plane.

**Results.**

**Structural properties of the heterostructures**

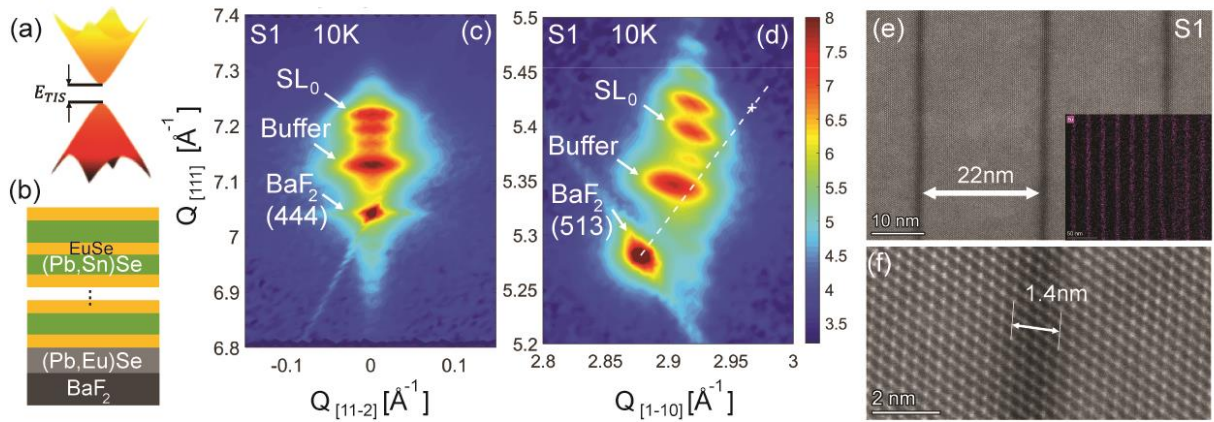

**Figure 1. Crystal structure of sample S1.** (a) Sketch of the massive Dirac dispersion of topological interface states after gap opening. (b) Diagram of the sample structure. The EuSe/$Pb_{1-x}Sn_xSe$/EuSe structure is repeated 5 or 10 times. (c) Reciprocal space map (RSM) taken on sample S1 showing the (444) Bragg peak of the $BaF_2$ substrate, the $Pb_{1-y}Eu_ySe$ buffer layer and the $Pb_{1-x}Sn_xse$/EuSe superlattice (SL). (d) Reciprocal space map of the (513) Brag reflection taken on S1. RSMs are taken at T=10K. The scale bar quantifies the intensity on a logarithmic scale. (e) Transmission electron microscopy image of a few wells from sample S1. Bottom right inset: Energy dispersive X-ray (EDX) spectroscopy map of the Eu distribution taken across all 10 periods of sample S1. (f) Zoom-in at the interface between $Pb_{1-x}Sn_xSe$ (light) and the EuSe barrier (dark).

The Pb$_{1-x}$Sn$_x$Se/EuSe superlattices oriented in the (111) direction are grown on BaF$_2$ (111) substrates by MBE. First, a thick buffer layer of insulating Pb$_{0.84}$Eu$_{0.16}$Se (400-500nm) is grown to reduce the lattice mismatch between the Pb$_{1-x}$Sn$_x$Se wells (a=6.107-6.088Å) and the substrate (a=6.196Å). Above the buffer layer we grow a periodic stack of Pb$_{1-x}$Sn$_x$Se/EuSe multiquantum wells. We study 5 samples listed in table 1. Two samples N1 and N2 are dedicated to neutron reflectivity measurements. We focus on samples S1 and N1 in the manuscript.

X-ray diffraction (XRD) measurements carried out at T=10K, allow us to extract the strain state of S1 at temperatures of interest. In Fig. 1(c), a reciprocal space map (RSM) taken along (444) at T=10K shows the Bragg peaks of the substrate, the (Pb,Eu)Se buffer layer and the Pb$_{1-x}$Sn$_x$Se well for sample S1. Periodic peaks from the superlattice structure are also resolved indicating a highly coherent heterostructure. The patterns allow us to extract the superlattice period for each sample. This is consistently checked with transmission electron microscopy (TEM) measurements and X-ray reflectometry (XRR) (see supplementary note 1) to extract the well and EuSe thickness separately. The structural properties of the samples are shown in Table 1. In Fig. 1(d), an RSM along (513) allows us to compute the in-plane and out-of-plane lattice constant and extract information about the strain. The superlattice is under tensile strain ($\epsilon_\parallel \approx +0.4\%$) in the in-plane direction and compressive strain in the [111] direction.

Transmission electron microscopy images confirm the observations of XRD. Fig. 1(e) shows a well-aligned stack of Pb$_{1-x}$Sn$_x$Se and EuSe probed using TEM in sample S1. The inset of this image displays an elemental mapping of the Eu atom using energy dispersive X-ray spectroscopy (EDX) that demonstrates the periodic repetition of the EuSe layers. A zoom-in figure (Fig. 1(f)) shows an interface between the well and the barrier demonstrating a short-range roughness of about a monolayer. We highlight that samples with thicker EuSe layers (>3nm) yielded rougher interfaces, therefore we restrict this analysis to thinner layers. Additional XRD, XRR and TEM measurements are included in supplementary note 1. The large gap of EuSe (>1eV) prohibits interactions between neighboring quantum wells, so the heterostructure can be safely considered as a multi-quantum well (see supplementary note 2).[34]

**Magnetooptical measurements**

We employ magneto-optical infrared spectroscopy to probe the energy gap of the TIS in proximity to EuSe. Experiments are performed in applied magnetic fields up to 17.5T using both transmission and reflection measurements (Fig. 2(a)) and up to 35T using absorption. The transmission spectra revealed a very weak signal between 25 and 72 meV corresponding to the reststrahlen band of the BaF$_2$ substrate. Therefore, we employed the reflection geometry to study the infrared signal from the heterostructure in this spectral range. In the presence of a magnetic field, infrared excites transitions between various Landau levels (LLs) as illustrated in Fig. 2(b). The minima in the normalized transmission spectra T(B)/T(B=0) allow us to extract the energy of these transitions.

The experimental spectra obtained using transmission measurements are shown in Fig. 2(c,d) for S1. Owing to the high mobility of the TIS in $Pb_{1-x}Sn_xSe$, LL transitions that shift to higher energy as the field increases can be observed at fields as low as 3T. Some are marked with blue and gray dots to highlight the field dependence. Color coding will become clear later.

To obtain information about the transitions occurring in the reststrahlen of the substrate, we carried out reflectivity measurements in the far-infrared. The relative reflectance is shown in Fig. 2(e). Below 75meV, reflectance is strong, and a prominent minimum marked with a black square is observed. Other field-independent modulations of the reflectivity are ignored, since they cannot be related to LLs and are of no interest for our analysis.

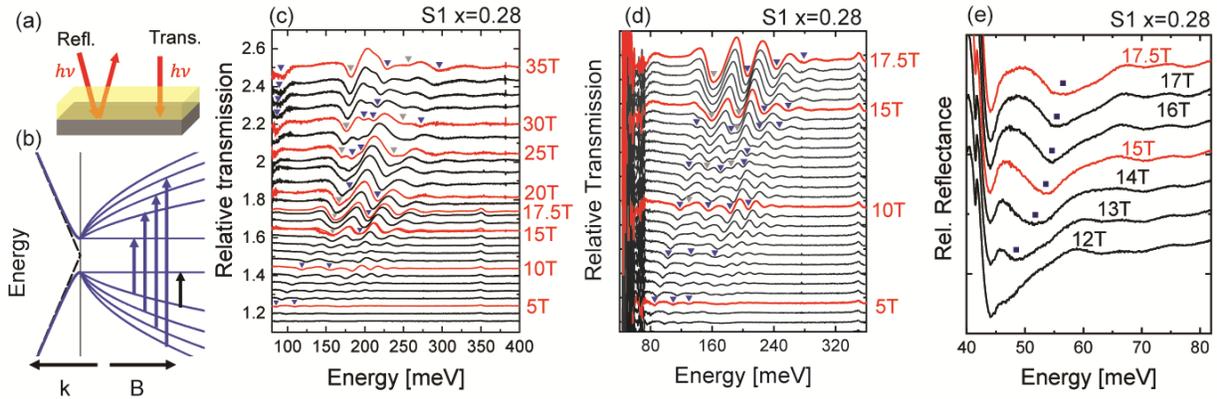

**Figure 2. Magnetooptical measurements taken on S1.** Landau level (LL) spectroscopy setup in the (a) transmission and reflection geometry. Reflectance is carried out in the reststrahlen band of the substrate. (b) Band dispersion and LLs of gapped TIS versus wavenumber k and versus magnetic field B. Blue arrows represent interband Landau level transitions and the black arrow is the cyclotron resonance (CR) (c) Relative magnetooptical transmission spectra (T(B)/T(0)) measured between 3T and 35T on S1. (d) Zoom-in showing measurements taken between 3 and 17.5T. (e) Relative reflectance spectra (T(B)/T(0)) measured between 12T and 17.5T on S1. All measurements are made at 4.5K. The cyclotron resonance is marked with the black square, the structure of peaks and troughs near 40meV is due to a $BaF_2$ phonon. The red curves highlight specific values of magnetic field for ease of reading.

We identify all transitions observed as minima in Fig. 2(c,d) by points presented in Fig. 3(a). Those transitions are shown in the empirical LL graphs plotted in Fig. 3(b). By modelling these LL transitions using a massive Dirac Hamiltonian[32] we can determine their origin. The transitions marked in blue in Fig. 2(c,d) and Fig. 3(a) are due to interband transitions between the conduction and valence LLs of the TIS. The intraband band transition between the $0^{th}$ and the $1^{st}$ LL is shown by the black square. Those marked in gray are associated with interband transitions from the LLs of trivial quantum well subbands. We make this attribution following analysis using the semi-empirical model discussed next.

Fig. 3(a) shows a curve fit carried out using the following relation which describes interband magnetooptical transitions for massive Dirac fermions given dipole selection rules [42,43] (see supplementary note 2):

$$\epsilon_n^{E,i} - \epsilon_{n\pm1}^{H,i} = \sqrt{\Delta_i^2 + 2ev^2\hbar nB} + \sqrt{\Delta_i^2 + 2ev^2\hbar(n\pm1)B}$$

Here $\Delta_i$ the band edge position of each subband, $i$ is the subband index ($i=1$ for the TIS), E/H denote the conduction and valence subbands respectively, $v$ is the band velocity taken to be the same for all subbands, $n$ is the Landau index, and $B$ is the magnetic field. $\epsilon_n^{E,i}$ thus denotes the energy of the $n^{th}$ LL of the $i^{th}$ electron subband. A k.p formalism using the envelope function scheme allows us to derive the band dispersion to show that it indeed satisfies this quasi-ideal massive Dirac model (see supplementary notes 2 and 3).[32,44] The solid blue lines in Fig. 3(a) represent transitions obtained from Eq. (1) for the TIS and the gray lines represent those from trivial QW subbands. Their extrapolation to B=0 yields the gap between each pair of conduction and valence subbands with the same index $i$ (i.e. $2\Delta_i$ the gap between $E_i$ and $H_i$). $2\Delta_1$ is $E_{TIS}$. It is zero for a thick quantum well and if time-reversal symmetry is preserved. The corresponding LLs for all subbands up to i=3 are shown in Fig. 3(b) with the transitions shown as arrows.

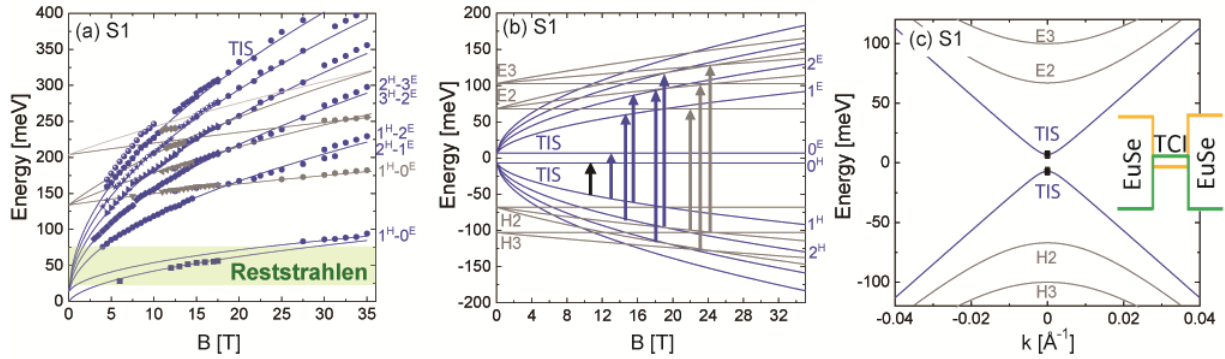

**Figure 3. Landau levels of sample S1 EuSe-Pb$_{0.72}$Sn$_{0.28}$Se-EuSe**. (a) Fan chart extracted from the experimental measurements and compared to the empirical model. The solid lines are the calculated transitions. They follow the same color scheme. (b) Landau levels of Pb$_{0.72}$Sn$_{0.28}$Se-EuSe utilized to compute the transitions shown in (a). Blue points and arrows represent the interband transitions of the topological interface states (TIS), the gray points and arrows represent those of QW subbands $E_i$ and $H_i$ (i>2) and the arrow represents the cyclotron resonance of the TIS. The numbers on the left (n$^{E/VH}$) label the conduction and valence Landau levels of the TIS. The dashed red line marks the position of the Fermi energy. (c) Band dispersion of Pb$_{0.72}$Sn$_{0.28}$Se-EuSe obtained from the experiment. The uncertainty on the TIS energy gap is shown at the band edge. The inset shows the band alignment and the topological states peaked at the interface.

The cyclotron resonance (CR) corresponds to the intraband transition between the valence n=0 and n=1 levels of the TIS, whose energy is given by:

$$\epsilon_{CR} = \sqrt{\Delta_1^2 + 2ev^2\hbar nB} - \Delta_1$$

This transition is shown as the black line in Fig. 3(a) and the black arrow in Fig. 3(b). The fit to the interband transitions and the CR in Fig. 3(a) yield the following gaps: the TIS gap $E_{TIS} = 2\Delta_1 = 14 \pm 6 meV$, the gap between E$_2$ and H$_2$ $2\Delta_2 = 68 \pm 2 meV$, and between E$_3$ and H$_3$ $2\Delta_2 =$

$204 \pm 2 meV$. The Fermi energy can also be determined from to be 40meV from the field at which the 3-4 transition disappears as the n=3 LL crosses the Fermi level (Fig. 3(a,b)). This experimental measurement allows us to reconstruct the band dispersion (in the QW plane) shown in Fig. 3(c). Despite the presence of the EuSe layer, the TIS gap cannot exceed 20meV within one standard deviation. The fit also yields the velocity $v = 4.3 \pm 0.05 \times 10^5 m/s$ ($2.8 eV \cdot Å$) for both S1 and S2, in agreement with previous work.[45,46,32] Since we observe several transitions between 3T and 35T, our uncertainty on $v$ and $\Delta_i$ is small. Data taken on sample S2 is presented in supplementary note 4 and yields similar but less precise results as its mobility was lower.

The experimental results for both samples S1 and S2 are shown in Table 2. The extracted energy gaps are compared to a theoretical calculation carried out to determine the impact of quantum confinement and strain on the trivial and topological states (supplementary notes 2 and 3). They utilizes the band alignment shown in Fig. 3(c). The resulting gaps are also shown in Table 2. The TIS gap from the experiment is only slightly larger than the theoretical value calculated without including magnetic exchange. The calculated value after the inclusion of the impact of strain is within experimental error. Strain reduces the topological bulk gap of the $Pb_{0.72}Sn_{0.28}Se$, enhancing the hybridization between top and bottom TISs.[47,48,49] For this reason, it is found to enhance $E_{TIS}$. The experimental $E_{TIS}$ is also equal within error to the confinement gap found in $Pb_{1-x}Sn_xSe$/EuSe multiquantum wells in previous works[32].

**Temperature dependence.**

Additional temperature dependent magnetooptical spectroscopy measurements are carried out to further corroborate the origin of the energy gap in S1. Spectra taken at various temperatures for B=13T are shown in Fig. 4(a). The transition involving the $N=1^H$ and $N=2^E$ transition of the interface states is seen to vary in energy versus temperature, starting at 64K. At 1.6, 4.2 and 6K, the magnetooptical spectra are nearly-identical. The energy dependence at high temperatures is analyzed by fitting to the model discussed above, to extract $E_{TIS}$. The fan-charts shown in Fig. 4(b-e) demonstrate the changing energy gap, extracted by extrapolating the Landau levels to zero field using the massive Dirac model. Both $E_{TIS}$ and $2\Delta_2$ are extract and plotted in Fig. 4(f). $E_{TIS}$ is constant at the lowest temperatures 1.6K to 6K, where we expect the magnetic ordering transition of EuSe to take place. It increases between 64K and 120K. This increase can be fully understood as the result of the increasing penetration depth of surface states as a function of temperature,[32] caused by the decreasing bulk energy gap of $Pb_{1-x}Sn_xSe$.[37,50] When this penetration depth increases, the hybridization between the top and bottom surfaces is enhanced and causes an enhanced gapping of their Dirac spectrum. These measurements suggest that the magnetic gap induced by proximity with EuSe is much smaller than the confinement gap at low temperatures.

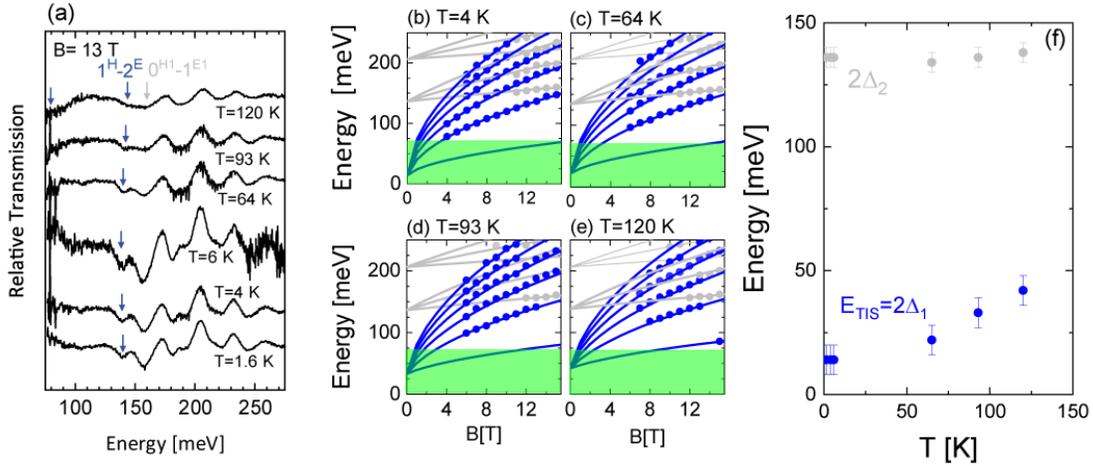

**Figure 4. Temperature dependent magnetooptical spectra of S1** (a) Magnetooptical spectra at B=13T measured on S1 at various temperatures. (b-e) Landau fan-charts extracted from the experimental measurements and compared to the empirical model. Blue points represent the interband transitions of the topological interface states (TIS), the gray points represent those of the first trivial quantum well state. Solid lines represent the calculated transitions. (f) The energy gap of the TIS ($E_{TIS}$) and the gap of the first quantum well subband ($2\Delta_2$) extracted from the temperature dependence analysis. The error bars in (f) present the uncertainty on the gap as a fit parameter from the curves shown in (b-e)

**Magnetic measurements.**

We have lastly performed SQUID magnetization measurements to identify the magnetic phases of our EuSe barriers. Fig. 5(a) shows the magnetic moment versus temperature with B=100G applied parallel to the sample surface. The curve consists of two main contributions: the paramagnetic signal (PM) of the $Pb_{1-x}Eu_xSe$ buffer layer, which increases as the temperature decreases following the Curie law and the antiferromagnetic (AFM) signal from the EuSe barriers, which yields a peak at the Néel temperature. All curves shown in Fig. 5(a) exhibit such a peak, indicating AFM order from the EuSe barriers. The extracted Néel temperature $T_N$ is plotted in Fig. 5(b). It increases with Sn concentration. This is likely due to the strain caused by a change in the lattice parameter of the $Pb_{1-x}Sn_xSe$ layer with increasing x.[51] The field-dependent magnetization was also measured and shown in Fig. 5(c) for the three samples considered here at 2K. Based on the estimated thickness we have calculated the average magnetic moment $\overline{m}_{Eu}$ per $Eu^{2+}$ in Fig. 5(c) after the correction for diamagnetism of the $BaF_2$ substrate and the PM of the buffer layer (see supplementary note 5). The fine structure of the magnetization versus magnetic field – reflecting transitions from an AFM ground state to a ferrimagnetic phase and finally to a saturated state – is only visible for samples with a thick EuSe layer ($\geq 4nm$). But, the magnetic moment saturates to a value close to the expected saturation from $Eu^{2+}$ ions, $7\mu_B/Eu^{2+}$.[52,53] Thus, for magnetic fields at which optical measurements are carried out, the moment of the EuSe layers can be considered saturated out-of-plane along the growth axis of the layer, and the extrapolation of the LLs to B=0 should yield $E_{TIS}$ in the presence of a ferromagnet in proximity.

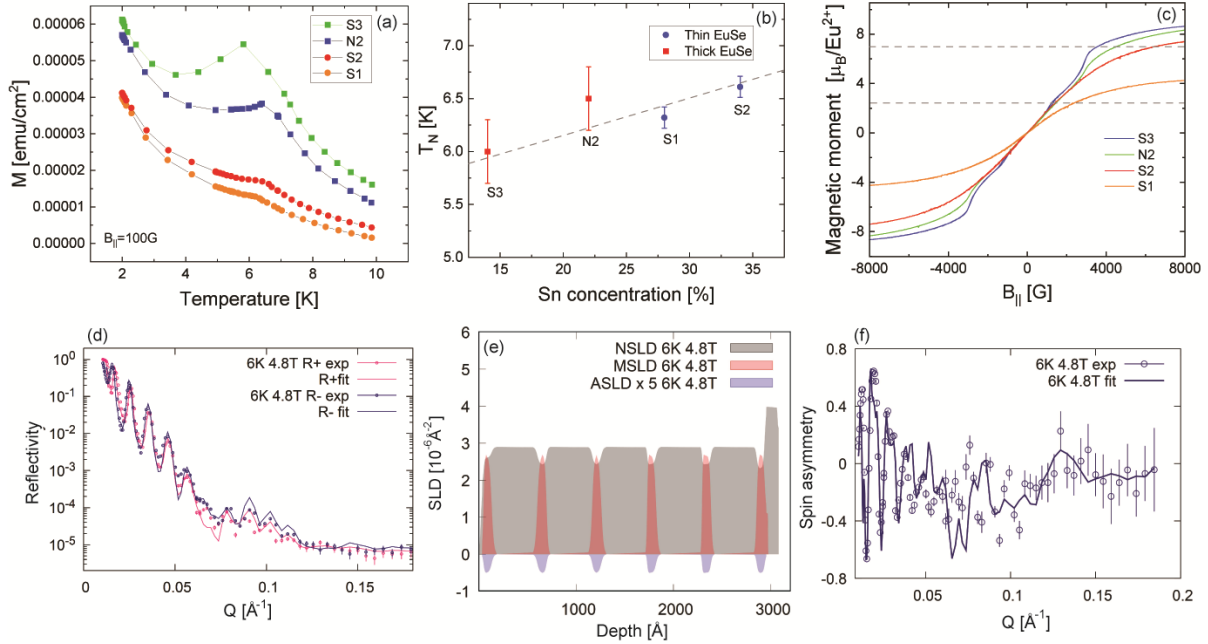

**Figure 5. Magnetic properties and polarized neutron reflectivity measurements.** (a) Magnetization versus temperature at 0.01T for 4 samples with different Sn concentrations. (b) The Neel temperature extracted from (a) for each sample plotted versus Sn concentration. (c) Magnetization versus magnetic field for samples with different Sn content. The dashed lines indicate the expected magnetic moment per $Eu^{2+}$ in the ferrimagnetic and ferromagnetic states of EuSe. (d) Measured (symbols) and fitted (solid lines) neutron reflectivity curves for spin-up ($R_+$) and spin-down ($R_-$) neutron spin-states as a function of wavevector transfer Q = 4πsin(θ)/λ, where θ is the incident angle and λ is the neutron wavelength. (e) Polarized neutron reflectometry nuclear (NSLD, in grey), magnetic (MSLD, in pink) and absorption (ASLD, in blue) scattering length density profiles, measured 6 K with an external in-plane magnetic field of 4.5 T and presented as a function of depth. The finite ASLD is due to the presence of absorbing Eu atoms.[54] (f) The spin asymmetry ratio, $SA = (R_+ - R_-)/(R_+ + R_-)$, obtained from the experimental and fitted reflectivity data in (d).

To probe the depth profile of the magnetism at buried interfaces, we carried out PNR measurements on two samples (N1 and N2) at Oak Ridge National Lab. Data for N1 is shown in Figure 5 and N2 is discussed in supplementary note 6. The PNR measurements yield the spin resolved reflectivities R+ and R– as a function of the wavevector transfer Q at a fixed magnetic field (Fig. 5(d)). Superscripts plus (or minus) denote measurements with neutrons with spin parallel (or antiparallel) to the direction of the applied magnetic field. The depth profiles of the nuclear and magnetic scattering length densities (NSLD and MSLD) correspond to the depth profile of the chemical and in-plane magnetization vector distributions, respectively. Both are extracted using the fit shown in Fig. 5(d) and plotted in Fig. 5(e). Fig. 5(f) shows the spin asymmetry ratio $SA = (R_+ - R_-)/(R_+ + R_-)$ obtained from the experimental and fitted curves at ±4.85T. The SA signal evidences the presence of a depth dependent magnetic moment.

The simulated depth profile in Fig. 5(e) and the SA in Fig. 5(f) confirm that we successfully obtain the intended periodic repetition of magnetic layers in proximity to every $Pb_{1-x}Sn_xSe$ layer. The peak MSLD signal is nearly constant in each of the 7 layers and converts to a value close to $5.5\mu_B/Eu^{2+}$ slightly lower than the value recovered from magnetometry. Despite the interface being atomically sharp at the nm scale, some roughness can be resolved in TEM measurements at the $\mu m$ scale. This can cause a slight reduction of the magnetic moment probed by PNR.

**Discussion.**

We have thus extracted $E_{TIS}$ for topological states in proximity to a magnetic insulator using Landau level spectroscopy. This is made possible by the synthesis of $Pb_{1-x}Sn_xSe$-EuSe multiquantum wells that achieve record mobility close to 17000cm$^2$/Vs (see supplementary note 7). Using the known bulk band parameters of $Pb_{1-x}Sn_xSe$ with x=0.28[37] and including the impact of quantum confinement and in-plane tensile strain on the energy levels (see supplementary note 3) we have reproduced the experimental energy gap without needing to include magnetic exchange interactions. The temperature dependence of $E_{TIS}$ agrees with this assumption. The size of the magnetic exchange gap, is thus within the error bars shown in Fig. 3(c) and Fig. 4(f). It is small and not conclusively detectable despite our detection of magnetic order at the interface. By comparing the top bound of our experimental gap from S1 to the theoretically computed gap, we can reliable set the upper bound of the magnetic gap to be 10meV.

We highlight here that crystalline symmetry breaking, resulting from a lattice distortion has also been shown to gap the (001) topological surface states of TCIs in previous works, even in the absence of magnetism.[55,56,45] However, the Dirac surface states at (111) surfaces have been found to be gapless by ARPES measurements down to low temperatures.[46,57]. The rhombic strain present in our layers, also should not break mirror symmetry at the (111) surface, also ruling out its role in generting a non-magnetic gap. Additionally, step edges in TCIs also on (001) surfaces have been shown to alter the electronic structure and to generate one-dimensional flat bands.[58] Such effects cannot account for our recovered energy gap, they are however interesting to consider in future studies on the (001) surfaces of TCIs, even in the presence of magnetism. Lastly, roughness and interface modification was shown to impact the momentum space position and splitting of Dirac cones on the (001) surface.[59] While it has not been studied for the case of (111) surfaces, this findings should not influence our conclusion since the Dirac nodes on the (111) surface are pinned to specific momenta at $\bar{\Gamma}$ and $\bar{M}$, unlike those on the (001) surfaces, which occur along $\bar{\Gamma} - \bar{X}$ lines.[57]

While the magnetic proximity induced gap of the TIS has never been measured, it has been computed for the $Bi_2Se_3$/EuS interface. Previous ab-initio simulations on this type of structure have yielded a small magnetic Dirac gap close to 3meV for the interface, and 9meV for the top surface.[4] The small gap obtained for the EuS-$Bi_2Se_3$ interface is associated to the highly localized nature of the f-orbital in Eu. At that interface, evidence of charge transfer was observed in experiments[60] and ab-initio calculations[61]. In our system, charge transfer into the MI is unlikely since the two materials should have a type-I band alignment and since the structure inherently preserves inversion symmetry. But, the scenario of orbital hybridization with the localized 4f-level could also hold which can explain our result that the proximity induced gap is small[4]. This allows us to draw an important conclusion. The magnetic proximity effect yields a narrow energy-gap at the interface between a TCI and rare earth containing insulators such as EuSe, which is not favorable for the realization of a stable QAHE. However, A narrow gap is advantageous for spintronic devices, since it is expected to preserve spin-momentum locking at the Fermi energy far-away from the band edges.[23]

**Methods**

**MBE synthesis.** The heterostructures are grown by molecular beam epitaxy on $BaF_2$(111) substrates. A compounds source of PbSe is utilized along with elemental sources of the following elements: Sn, Se, and Eu. A thick buffer layer of (Pb,Eu)Se (400-500nm) is initially grown to minimize dislocations in the heterostructures of interest. The buffer layer is followed by a EuSe/$Pb_{1-x}Sn_xSe$ heterostructure with either 5 or 10 periods. The thickness of the well is denoted by $d$ and that of the barrier by $d_B$. They are controlled by the growth time. Thickness of the EuSe barrier differs for each sample but is kept larger than or equal to 1.1nm (3ML). We chose to synthesize the Se-based TCIs for which Eu intermixing is known to be minimal.[62] The substrate temperature is fixed at 370°C during the growth and the Sn content x is varied by adjusting the relative flux between the PbSe cell and a Sn cell. A constant Se over-pressure is maintained throughout the growth. The well composition in each case is determined either by growing a control bulk epilayer of $Pb_{1-x}Sn_xSe$ (>200nm) under the same conditions and during the same run as the heterostructures and extracting its composition using X-ray diffraction, or by carrying out energy dispersive X-ray spectroscopy.

**Magnetooptical measurements.** Magneto-infrared experiments were performed using a commercial FT-IR spectrometer coupled with the vertical bore superconducting magnet reaching 17.5T and a resistive magnetic reaching 35T at the National High Magnetic Fields Lab. The IR radiation propagates inside an evacuated metal tube from the spectrometer to the top of the magnet, whereas a brass light-pipe is used to guide the IR radiation down to the sample space. A parabolic cone was used to collect IR beam at the sample while two mirrors reflected the beam back up towards to the Si composite bolometer mounted just a short distance above the sample space. The sample is cooled down to 5 K using low pressure He exchange gas. The IR spectra were recorded in the mid-IR and far-IR range at fixed magnetic field, while field was stepped between 0 and 35T with increment of 0.5T. All measurements are carried out in the Faraday geometry with the incident propagating perpendicular to the surface and along the direction of the applied field. The reported spectra are obtained by dividing the signal at finite magnetic field by the zero-field spectrum.

**Polarized Neutron Reflectometry.** Polarized neutron reflectometry experiments were performed on the Magnetism Reflectometer at the Spallation Neutron Source at Oak Ridge National Laboratory,[63] using neutrons with wavelengths $\lambda$ in a band of 2–8 Å and a high polarization of 98.5–99%. PNR is a highly penetrating depth-sensitive technique that can probe the chemical and magnetic depth profiles of materials with a resolution of 0.5 nm. The depth profiles of the nuclear and magnetic scattering length densities (NSLD and MSLD) correspond to the depth profile of the chemical and in-plane magnetization vector distributions, respectively[64,65,66,67,68] Measurements were conducted in a closed cycle refrigerator equipped with a 5 T cryomagnet. Using the time-of-flight method, a collimated polychromatic beam of polarized neutrons with the wavelength band Δλ impinges on the film at a grazing angle $\vartheta$, interacting with atomic nuclei and the spins of unpaired electrons. The reflected intensity $R_+$ and $R_-$ are measured as a function of wave vector momentum, $Q = 4\pi\sin(\theta)/\lambda$, with the neutron spin parallel (+) or antiparallel (−) to the applied field. To separate the nuclear from the magnetic scattering, the spin asymmetry ratio SA = $(R_+ - R_-)/(R_+ + R_-)$ is calculated, for which SA = 0 designating no magnetic moment in the system.

**X-ray diffraction**. X-ray diffraction measurements are carried out at room temperature in a Bruker D8 Discover diffractometer equipped with Cu-Kα-source. Low temperature X-ray diffraction measurements are carried out at beamline 33ID-D/E at the Advanced Photon Source at Argonne National lab using a wavelength of 0.61992Å.

**SQUID magnetometry.** SQUID magnetometry is carried out in a Quantum Design MPMS, down to 4.2 K at various magnetic fields up to 7 $T$. The field is applied parallel to the sample plane. The diamagnetism of the substrate measured at 2K is subtracted from magnetization versus field measurements. The paramagnetism of the buffer layer is accounted for by a saturating Brillouin function and can be differentiated from the intrinsic magnetic properties of the EuSe layer using the method discussed in the supplementary note 5.

**Transmission electron microscopy.** High-resolution cross-sectional TEM images were acquired using a double tilt holder and probe corrected Spectra 30-300 transmission electron microscope (Thermo Fisher Scientific, USA) equipped with a field emission gun, operated at 300 kV. STEM images were acquired using Panther STEM detector (Thermo Fisher Scientific, USA) in high-angle, annular dark field mode (HAADF) and bright field mode (BF). For compositional analysis, energy-dispersive X-ray spectroscopy (EDX) maps were obtained in STEM mode using the Super-X EDX system (Thermo Fisher Scientific, USA) equipped with 4 windowless silicon drift detectors. TEM samples were prepared by focused ion beam etching using the standard lift-out technique.

**Theoretical Model.** The band structure of the nontrivial $Pb_{1-x}Sn_xSe$ heterostructures can be theoretically calculated using an envelope function model developed in previous work (for details see supplementary note 1).[32,33] The model does not take magnetism into account and has been shown to agree with experiments when the barrier is a simple insulator with a large energy gap (1eV) without magnetic order.[32] The inverted band structure in the non-trivial $Pb_{1-x}Sn_xSe$ will lead to a confined topological interface state (TIS). The bulk bands of the system also yield trivial quantum well subbands. The dispersion of these bands is calculated by an empirical 4-band massive Dirac model that includes a small parabolic correction:

$$\epsilon(k) = \pm\sqrt{\left(\Delta_i + \frac{\hbar^2 k^2}{2\widetilde{m}}\right)^2 + \hbar^2 v^2 k^2}$$

Where $\Delta_i$ is the energy gap, $v$ is the band velocity equal for the bulk and the confined system and k is the wavevector in the quantum well plane. The LLs of each band can be computed in a similar way:

$$\epsilon_{n,\uparrow}^{E/H}(B) = +\hbar\widetilde{\omega} \pm \sqrt{(\Delta_i + \hbar n\widetilde{\omega})^2 + 2ev^2\hbar nB}$$

$$\epsilon_{n,\downarrow}^{E/H}(B) = -\hbar\widetilde{\omega} \pm \sqrt{(\Delta_i + \hbar n\widetilde{\omega})^2 + 2ev^2\hbar nB}$$

$\widetilde{\omega} = eB/\widetilde{m}$ is the far-band correction to the effective mass $\widetilde{m}$. It is generally yielding a massive Dirac model. The arrows indicate the pseudo-spin direction. E/H label the conduction and valence levels.

In the magneto-optical measurment in the Faraday geometry, the selection rules require that electrons can only transition from n[H] (the nth LL in valence band) to $(n \pm 1)^E$, and vice versa. Eq. (1) is derived with this in mind. More details concerning the model are given in supplementary note 2.

**Data availabity**

The data that support the findings of this study are available from the corresponding author upon reasonable request.

**Code availabity**

Code that supports the findings of this study is available from the corresponding author upon reasonable request.

**Acknowledgements**

Work support by NSF-DMR-1905277. A portion of this work was performed at the National High Magnetic Field Laboratory, which is supported by National Science Foundation Cooperative Agreements No. DMR-1644779, DMR-2128556, and the State of Florida. This research used resources at the Spallation Neutron Source, a Department of Energy Office of Science User Facility operated by the Oak Ridge National Laboratory. XRR measurements were conducted at the Center for Nanophase Materials Sciences (CNMS), which is a DOE Office of Science User Facility. We also acknowledge support from the Notre Dame Integrated Imaging Facility. This research used resources of the Advanced Photon Source, a U.S. Department of Energy (DOE) Office of Science User Facility, operated for the DOE Office of Science by Argonne National Laboratory under Contract No. DE-AC02–06CH11357.

**Author contributions**

JW, XL and BAA conceived the project. JW, MO and SKB carried out the magnetooptical experiments with input form BAA and DS. JW, HT and SKB analyzed the magnetooptical data. TW and JW carried out the k.p calculations. JBO, LAV and YG carried out temperature dependent magnetooptical measurements and analyzed them. JW, ZZ and BAA carried out X-ray diffraction measurements. MZ and TO carried out TEM measurements. VL, HA and JK carried out and analyzed neutron reflectivity and X-ray reflectivity measurements. XL synthesized the samples. JW, VL, XL and BAA interpreted the data from various experiments and wrote the manuscript.

**Ethics declarations**

The authors declare no competing interests.

| Sample | Sn concentration x | $Pb_{1-x}Sn_xSe$ Well thickness(nm) | EuSe Barrier thickness(nm) | No. of periods |
|---|---|---|---|---|
| S1 | 0.28±0.01 by EDX | 22(±1) | 1.41±0.35 (4±1ML) | 10 |
| S2 | 0.25±0.03 by EDX | 25(±2) | 2.4 | 5 |
| S3 | 0.14 by XRD | 32 | 6 | 5 |
| N1 | 0.10 by XRD | 48 | 11 | 5 |
| N2 | 0.22 by XRD | 35 | 4±2, varies by layer | 5 |

**Table 1. Sample list.** The Sn concentration is determined using energy dispersive X-ray spectroscopy (EDX) performed during Transmission electron microscopy (TEM) and X-ray diffraction (XRD) on bulk control samples. The well and barrier thickness are extracted from a period determined by XRD of the superlattice and cross-sectional TEM measurements. The uncertainty accounts for the interface roughness. XRD patterns for all samples are shown in Supplementary note 1. ML: monolayer.

| Sample ID | $E_{TIS} = 2\Delta_1$ (meV) | E2-H2 gap $2\Delta_2$ | E3-H3 gap $2\Delta_3$ |
|---|---|---|---|
| S1 – Exp. | 14±6 | 136(±4) meV | 206 meV |
| S1 – Cal. | 7 | 129 meV | 204 meV |
| S1 – Cal. with strain | 10 | 122 meV | 198 meV |
| S2 – Exp. | 20±10 | 118±8 meV | 200 meV |
| S2 – Calc.* | 7±4 | 116meV | 183meV |
| S2 – Calc. with strain* | 18±6 | 106 meV | 179meV |

**Table 2. Experimental and theoretical energy gaps for S1 and S2.** The calculation utilizes an envelope function scheme implemented using the band alignment shown in the inset of Fig. 3(c) and discussed in the supplementary note 2. Magnetic exchange is neglected in these calculations.

The measurements for S2 are shown in supplementary note 4. *For S2, the uncertainty on the calculated gap includes the uncertainty on the composition of the well from energy dispersive X-ray spectroscopy.